\documentclass[journal, twocolumn]{IEEEtran}

\usepackage{amsmath,amssymb,amsfonts,amsthm}
\usepackage{textcomp}
\usepackage{color}
\usepackage[numbers,sort&compress]{natbib}
\usepackage{graphicx}
\usepackage{xspace}
\usepackage{bm}
\usepackage{array, makecell}
\usepackage{enumitem}
\usepackage[nolist,nohyperlinks]{acronym}

\def\BibTeX{{\rm B\kern-.05em{\sc i\kern-.025em b}\kern-.08em
    T\kern-.1667em\lower.7ex\hbox{E}\kern-.125emX}}
\makeatletter
\theoremstyle{plain}
\def\blfootnote{\xdef\@thefnmark{}\@footnotetext}
\makeatother

\newtheorem{definition}{Definition}
\newtheorem{lemma}{Lemma}
\newtheorem{corollary}{Corollary}

\begin{acronym}
\acro{SNR}{signal-to-noise ratio}
\acrodefplural{SNR}[SNRs]{signal-to-noise ratios}
\acro{RV}{random variable}
\acrodefplural{RV}[RVs]{random variables}
\acro{LoS}{line-of-sight}
\acrodefplural{LoS}{Line-of-sight}
\acro{IG}{inverse gamma}
\acro{PDF}{probability density function}
\acro{CDF}{cumulative distribution function}
\acrodefplural{CDF}[CDFs]{cumulative distribution functions}
\acro{MGF}{moment generating function}
\acro{GMGF}{generalized moment generating function}
\acrodefplural{GMGF}[GMGFs]{generalized moment generating functions}
\acro{TWDP}{two-wave with diffuse power}
\acro{MC}{Monte Carlo}
\acro{FTR}{Fluctuating Two-Ray}
\acro{URLLC}{ultra-reliable low-latency communications}
\acro{UAV}{unmanned aerial vehicle}
\acro{WOC}{wireless optical communications}
\end{acronym}

\newcommand{\MATLAB}{\textsc{Matlab}\xspace}
\newcommand{\MATHE}{\textsc{Mathematica}\xspace}
\DeclareMathOperator*{\argmin}{argmin}

\begin{document}


\title{Composite Fading Models based on Inverse \\Gamma Shadowing: Theory and Validation}
\author{Pablo Ram\'irez-Espinosa and F. Javier L\'opez-Mart\'inez}

\maketitle

\blfootnote{\noindent Manuscript received March xx, 2020; revised XXX. This work has been funded by the Spanish Government and the European Fund for Regional Development FEDER (project TEC2017-87913-R), by Junta de Andalucia (project P18-RT-3175, TETRA5G) and by University of M\'alaga. The review of this paper was coordinated by XXXX. A preliminary version of this paper was presented at IEEE Globecom 2019 \cite{Pablo2019}.}

\blfootnote{\noindent P. Ram\'irez-Espinosa is with the Connectivity Section,  Department  of  Electronic  Systems, Aalborg University, Aalborg {\O}st 9220, Denmark (e-mail: $\rm pres@es.aau.dk$).}

\blfootnote{\noindent F. J. L\'opez-Mart\'inez is with Departmento de Ingenieria  de Comunicaciones, Universidad de Malaga - Campus de Excelencia Internacional Andalucia Tech., Malaga 29071, Spain (e-mail: $\rm fjlopezm@ic.uma.es$).}

\blfootnote{This work has been submitted to the IEEE for publication. Copyright may be transferred without notice, after which this version may no longer be accesible.}

\begin{abstract}
We introduce a general approach to characterize composite fading models based on inverse gamma (IG) shadowing. We first determine to what extent the IG distribution is an adequate choice for modeling shadow fading, by means of a comprehensive test with field measurements and other distributions conventionally used for this purpose. Then, we prove that the probability density function and cumulative distribution function of any IG-based composite fading model are directly expressed in terms of a Laplace-domain statistic of the underlying fast fading model and, in some relevant cases, as a mixture of well-known state-of-the-art distributions. Also, exact and asymptotic expressions for the outage probability are provided, which are valid for any choice of baseline fading distribution. Finally, we exemplify our approach by presenting several application examples for IG-based composite fading models, for which their statistical characterization is directly obtained in a simple form.
\end{abstract}


\section{Introduction}

In wireless channels, the random fluctuations affecting  the radio signals have been classically divided into two types: fast fading, as a result of the multipath propagation, and shadow fading or shadowing, which is caused by the presence of large objects like trees or buildings. Aiming to study and improve the performance of wireless communication systems, considerable efforts have been devoted to the characterization of these two effects which, in many cases, are analyzed separately. Thus, several models are used to describe the statistical behavior of fast fading, including both the classical ones such as Rayleigh, Rice, Hoyt (Nakagami-$q$) and Nakagami-$m$ \cite{Alouini2005, Stuber02, Nakagami1960}, as well as generalized models that have gained considerable popularity in the last years \cite{Durgin2002,Yacoub2007, Fraidenraich2006}. With respect to shadow fading, the lognormal distribution is widely accepted as the most usual choice \cite{Stuber02, Hashemi1993}, supported by empirical verification.

Although fast fading and shadowing occur simultaneously in practice (at different time scales), they are often characterized as different phenomena and their analyses are carried out separately. Due to the inherent connection between fast and shadow fading, composite fading models arose to characterize the combined impact of these two effects, classically as the superposition of lognormal shadowing and some fading distribution. In the literature, there are two different ways to incorporate the effect of shadowing on the top of fading: (\emph{i}) multiplicative shadowing, in which both the specular and diffusely scattered fading components are shadowed \cite{Abdi1998,Yoo2018}, and (\emph{ii}) \ac{LoS} shadowing, where shadowing only affects the specular component \cite{Abdi2003}.

Examples of the latter type are the Rician shadowed \cite{Abdi2003}, the $\kappa$-$\mu$ shadowed \cite{Paris2014,Ramirez2018} and the \ac{FTR} \cite{Romero2017} fading distributions. On the other hand, multiplicative shadowing models originally arose as a combination of lognormally-distributed shadowing and classical fading models like Rayleigh or Nakagami-$m$ \cite{Stuber02, Loo1985}. However, all these models inherit the complicated formulation of the lognormal distribution, considerably limiting their usefulness for further analytical calculations.

As an alternative to the lognormal distribution, the gamma distribution has been proposed in the literature, showing its suitability to model shadowing through goodness-of-fit tests \cite{Abdi1999, Abdi2011}. Thanks to its mathematical tractability, new composite models have emerged by substituting the complicated lognormal distribution by the gamma distribution, e.g., the $K$ distribution (Gamma/Rayleigh) \cite{Abdi1998} and its generalization (Gamma/Gamma) \cite{Bithas2006}, Gamma/Weibull \cite{Bithas2009}, Gamma/$\kappa$-$\mu$ and Gamma/$\eta$-$\mu$ \cite{Hmood2017}. More sophisticated models, which combine the aforementioned two types of shadowing, were also recently introduced. For instance, in \cite{Simmons2019} the \ac{LoS} fluctuation due to human-body shadowing is combined with multiplicative shadowing, whilst in \cite{Simmons2020} several double-shadowed models are derived from the $\kappa$-$\mu$ fading distribution.

A different option to model shadowing is explored in \cite{Karmeshu2007, Eltoft2005, Sofotasios2013}, where the inverse Gaussian distribution is proposed. This approach proves to be specially accurate to approximate the lognormal distribution when the variance of shadowing is large. Finally, in the recent years, the \ac{IG} distribution has started to be used to characterize shadowing, motivated by the fact that it admits a relatively simple mathematical formulation. Based on the \ac{IG} distribution, different composite models have been proposed in \cite{Yoo2017, Yoo2018}. Despite its recent popularity, a rigorous empirical validation to assess the adequacy of the \ac{IG} distribution to model shadow fading has not been performed in depth. To the best of the authors' knowledge, only very brief validations of the \ac{IG} shadowing are carried out in \cite{Bithas2018, Yoo2019b}, but the results are scarce to be considered an exhaustive proof. Empirical validation of \ac{IG}-based composite models have also been addressed in \cite{Yoo2017, Yoo2018, Bithas2020}, {but the authors did not check that the shadowing alone follows an \ac{IG} distribution, and only showed that the whole composite distribution (shadowing and fading) fits measured data in those specific scenarios. 

Despite the lack of a proper shadowing validation, the works in \cite{Yoo2017, Yoo2018, Bithas2020} show a recent interest in the \ac{IG} distribution to model shadowing. More specifically,  [29] demonstrates that composite models based on \ac{IG} shadowing are excellent candidates to model propagation in the context of \ac{UAV} communications, one of the most popular areas in the last years in wireless communications \cite{Zeng2017, Wu2018}. On a related note, composite models are of widespread use in \ac{WOC} to model turbulence-induced fading, where the \ac{IG} is a solid alternative to lognormal and Gamma distributions  \cite{Ammar2001, Peppas2020, Jurado2011}.}

Taking into account the interest in the \ac{IG} distribution as shadowing model and having in mind the extensive number of composite models available in the literature, in this work, we aim to find answer to two key questions: 
\begin{enumerate}[label=(\alph*)]
	\item \emph{Is the use of the \ac{IG} distribution to model shadowing supported by practical evidences?}
	\item \emph{Does the \ac{IG} distribution bring additional benefits to other shadowing models?}
\end{enumerate}

In order to answer the first one, we perform an extensive set of goodness-of-fit tests using empirical data measurements. Once this is accomplished, we present a general approach to the statistical characterization of composite fading channels with \ac{IG} shadowing. Specifically, the contributions of this work are summarized as follows:
\begin{itemize}
	\item Motivated by Abdi's work \cite{Abdi1999} where the Gamma distribution is proposed as an alternative to lognormal shadowing, we perform a thorough validation of the \ac{IG} distribution as an alternative to gamma, inverse Gaussian and lognormal distributions. To this end, we target a large number of scenarios, both indoor and outdoor with frequencies ranging from a hundreds of MHz to millimeter-waves. 
	\item We show that the \ac{PDF} and the \ac{CDF} of the composite fading distribution can be directly expressed in terms of a generalization of the \ac{MGF} of the fast fading model. This holds for \textit{any} arbitrary choice of fading distribution, and allows to use existing results in the literature for state-of-the-art fading models to fully characterize the statistics of their composite counterpart.
	\item When the underlying fading distribution admits a representation as a mixture of gamma distributions, then the composite fading model can be expressed as a mixture of Fisher-Snedecor $\mathcal{F}$-distributions \cite{Yoo2017}. This is often the case for several popular fading models in the literature, and thus the formulation of their composite counterparts becomes straightforward. 
	\item The outage probability analysis is carried out for composite models with \ac{IG} shadowing, proving that the composite version inherits the same diversity order of the underlying fading distribution.   
	\item Finally, we exemplify our mathematical framework through the introduction of two general families of \ac{IG}-based composite models, based on the $\kappa$-$\mu$ shadowed and \ac{TWDP} fading distributions, respectively.
\end{itemize}

The rest of the paper is organized as follows. In Section \ref{sec:IGvalidation} the use of the \ac{IG} distribution to model shadow fading is empirically validated. Section \ref{sec:PhysicalModel} presents the physical model for \ac{IG}-based composite models. In Section \ref{sec:Statistics}, we introduce our general approach to characterize composite fading models with \ac{IG} shadowing, and also analyze the particular cases corresponding to integer \ac{IG} shape parameter, and to the mixture of gammas representation of the fading model, respectively. Section \ref{sec:OutageAnalysis} leverages the mathematical framework to characterize the outage probability of \ac{IG}-based composite models. After that, the derived results are exemplified to illustrate how a composite version of the popular and notoriously unwieldy \ac{TWDP} fading model \cite{Durgin2002} can be attained. Finally, the main conclusions are drawn in Section \ref{sec:Conclusions}.

\section{Empirical validation of IG shadowing}
\label{sec:IGvalidation}


\subsection{Definitions of shadowing distributions}
\label{sec_IIA}

\begin{definition}[Lognormal distribution]
	\label{def:LN}
	Let $X$ be a \ac{RV} following a Gaussian distribution with mean $\mu$ and variance $\sigma^2$. Then, the \ac{RV} $Y = e^X$ is lognormally distributed with \ac{CDF} 
	\begin{equation}
		F_{Y}^{\rm L} (\mu,\sigma;y) = \frac{1}{2} + \frac{1}{2}{\rm erf}\left(\frac{{\rm ln}\,y - \mu}{\sqrt{2\sigma^2}}\right) \label{eq:LognormalCDF}
	\end{equation}	 
	where ${\rm erf}(\cdot)$ is the error function \citep[eq. (7.1.1)]{Abra72}.
\end{definition}

\begin{definition}[Gamma distribution]
	Let $Y$ be a \ac{RV} following a gamma distribution with shape parameter $k$ and $\mathbb{E}[Y] = \Omega$, with $\mathbb{E}[\cdot]$ the mathematical expectation. Then, the \ac{PDF} and \ac{CDF} of $Y$ are given by 
	\begin{align}
		f_Y^{\rm G}(k,\Omega;y) &= \frac{k^k}{\Gamma(k)\Omega^k}y^{k-1}e^{-ky/\Omega},\label{eq:GammaPDF} \\
		F_Y^{\rm G} (k,\Omega;y) &= \frac{1}{\Gamma(k)}\gamma (k, ky/\Omega), \label{eq:GammaCDF}
	\end{align}	 
	with $\Gamma (\cdot)$ and $\gamma(\cdot,\cdot)$ the gamma function and the lower incomplete gamma function, respectively \cite[eqs. (6.1.1) and (6.5.2)]{Abra72}.
\end{definition}

\begin{definition}[Inverse Gaussian distribution]
	\label{def:InvGauss}
	Let $Y$ be a \ac{RV} following an inverse Gaussian distribution with parameters $\mu_I$ and $\lambda$. Then, the \ac{CDF} of $Y$ is given by
 	\begin{align}
		F_Y^{\rm i\mathcal{G}} (\mu_I,\lambda;y) =& \frac{1}{2} + \frac{1}{2}{\rm erf}\left(\sqrt{\frac{\lambda}{2y}}\left(\frac{y}{\mu_I}-1\right)\right) + \exp\left(\frac{2\lambda}{\mu_I}\right) \notag \\ 
		&\times\left[\frac{1}{2} + \frac{1}{2}{\rm erf}\left(-\sqrt{\frac{\lambda}{2y}}\left(\frac{y}{\mu_I}+1\right)\right)\right]. \label{eq:InvGaussCDF}
	\end{align}	 
\end{definition}
 
\begin{definition}[Inverse gamma distribution]
    \label{def:InvGamma}
	Let $Y$ be a \ac{RV} following an \ac{IG} distribution with shape parameter $m$ and $\mathbb{E}[Y] = \Omega_i$. Then, the \ac{PDF} and \ac{CDF} of $Y$ are given by
 	\begin{align}
 		f_Y^{\rm IG}(m, \Omega_i; y) &= \frac{\Omega_i^m(m-1)^m}{\Gamma(m)} y^{-m-1}e^{-\Omega_i(m-1)/y}, \label{eq:InvGammaPDF} \\
		F_Y^{\rm IG} (m, \Omega_i; y) &= \frac{1}{\Gamma(m)}\Gamma (m, \Omega_i(m-1)/y), \label{eq:InvGammaCDF}
	\end{align}	 
	where $\Gamma(\cdot,\cdot)$ is the upper incomplete gamma function \cite[eq. (6.5.3)]{Abra72}.
\end{definition}

\subsection{Fitting to field measurements}

\label{sec:Fitting}

In order to validate the suitability of the \ac{IG} distribution to model shadowing, we here compare empirical \acp{CDF} obtained from data measurements in a number of scenarios with the different models defined in the previous subsection, which are those commonly used in the literature to characterize shadow fading, i.e., lognormal, gamma, inverse Gaussian and \ac{IG}. As a goodness-of-fit measurement, we use the Cramer-von Mises test for the comparison, which is a more powerful option than the well-known Kolmogorov-Smirnov test. That is, the probability of accepting the alternative hypothesis when the alternative hypothesis is
true is higher in the Cramer-von Mises test \cite{Stephens1974, Abdi2011}. It is defined as the integrated mean square error between the empirical \ac{CDF}, $\widehat{F}_\xi(t)$, and the theoretical one, $F_\xi(t)$, i.e.,
\begin{equation}
	\label{eq:CramerTest}
	\omega^2 = \int_{-\infty}^{\infty}\left|\widehat{F}_{\xi}(t) - F_{\xi}(t)\right|^2 dt.
\end{equation}

\begin{figure}[t]
	\centering
     \includegraphics[width=1\columnwidth]{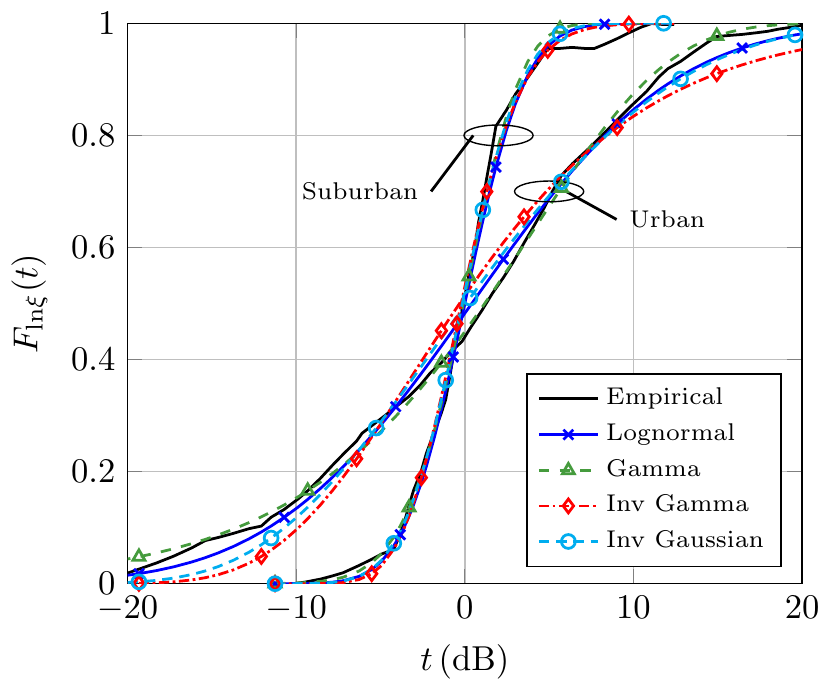}
         \caption{Shadowing CDF for data in \cite[Fig. 4]{Babiroli2017}, corresponding to urban and suburban scenarios at $169$ MHz. Parameters for urban scenario: lognormal ($\mu=0.05$, $\sigma=1.08$), gamma ($k = 1.15$, $\Omega = 1.57$), inverse gamma ($m = 1.18$, $\Omega_i = 4.60$) and inverse Gaussian ($\mu_i = 1.86$, $\lambda = 1.04$). Parameters for suburban scenario: lognormal ($\mu=0$, $\sigma=0.33$), gamma ($k = 10.15$, $\Omega = 1.02$), inverse gamma ($m = 9.82$, $\Omega_i = 1.05$) and inverse Gaussian ($\mu_i = 1.04$, $\lambda = 9.58$).}
      \label{fig:1}
\end{figure}

\begin{figure}[t]
	\centering
     \includegraphics[width=1\columnwidth]{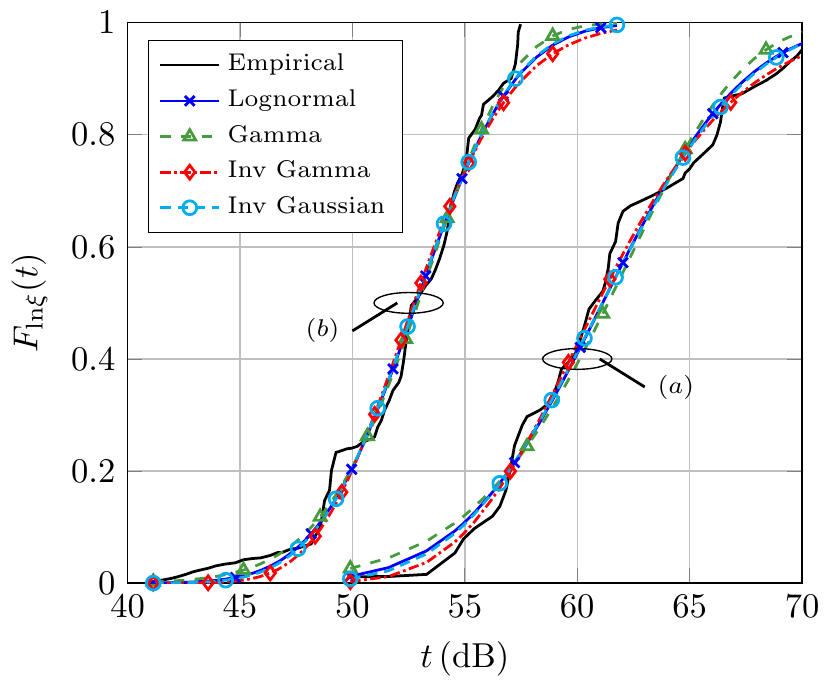}
         \caption{Shadowing CDF data in \cite[Fig. 1-2]{Abdi1998}, corresponding to suburban scenarios at $910.25$ MHz. Parameters for case $(a)$: lognormal ($\mu=7.04$, $\sigma=0.57$), gamma ($k = 3.45$, $\Omega = 1293$), inverse gamma ($m = 3.32$, $\Omega_i = 1433$) and inverse Gaussian ($\mu_i = 1345$, $\lambda = 3607$). Parameters for case $(b)$: lognormal ($\mu=6.08$, $\sigma=0.40$), gamma ($k = 6.80$, $\Omega = 465.4$), inverse gamma ($m = 6.54$, $\Omega_i=485.7$) and inverse Gaussian ($\mu_i = 473.4$, $\lambda = 2848$).}
      \label{fig:2}
\end{figure}

\begin{figure}[t]
	\centering
     \includegraphics[width=1\columnwidth]{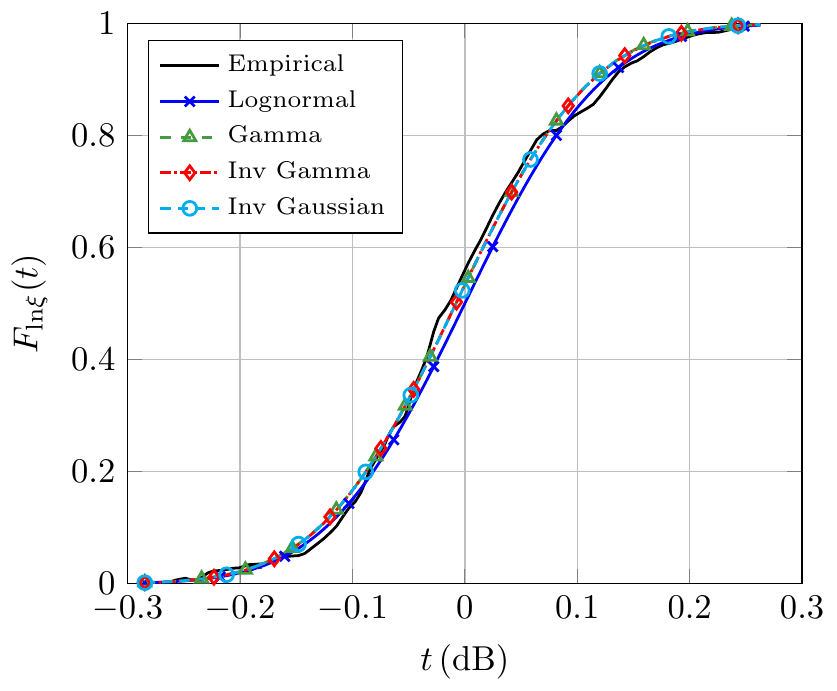}
         \caption{Shadowing CDF for data in \cite[Fig. 3b]{Ai2017}, corresponding to an indoor scenario at $26$ GHz. Parameters: lognormal ($\mu=0$, $\sigma=0.011$), gamma ($k = 8319$, $\Omega = 0.998$), inverse gamma ($m = 8316$, $\Omega_i=1$) and inverse Gaussian ($\mu_i = 1$, $\lambda = 8310$).}
      \label{fig:3}
\end{figure}

\begin{figure}[t]
	\centering
     \includegraphics[width=1\columnwidth]{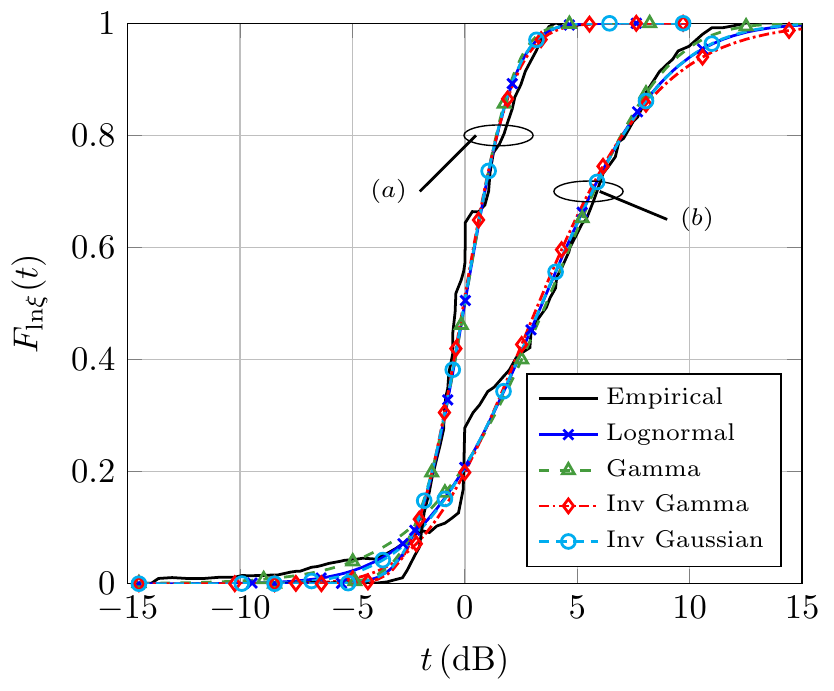}
         \caption{Shadowing CDF for data in \cite[Fig. 4b]{Zhu2015}, corresponding to an indoor scenario at $45$ GHz with two distinct types of antennas: horn antenna (case $(a)$) and open ended guide antenna (case $(b)$). Parameters for case $(a)$: lognormal ($\mu=0$, $\sigma=0.20$), gamma ($k = 26.49$, $\Omega = 1.01$), inverse gamma ($m = 26.17$, $\Omega_i=1.02$) and inverse Gaussian ($\mu_i = 1$, $\lambda = 25.74$). Parameters for case $(b)$: lognormal ($\mu=0.4$, $\sigma=0.49$), gamma ($k = 4.52$, $\Omega = 1.64$), inverse gamma ($m = 4.69$, $\Omega_i=1.72$) and inverse Gaussian ($\mu_i = 1.67$, $\lambda = 6.51$).}
      \label{fig:4}
\end{figure}

\begin{table*}[t]

\caption{Results of the Cram\'er-von Mises test, $\omega^2$, for data in Figs. \ref{fig:1}-\ref{fig:5}. Showed values are normalized by $10^{-3}$. Lower values are better.}
\label{Table1}
\centering
\setcellgapes{5pt}\makegapedcells
\begin{tabular}
{|c||c|c|c|c|c|c|c|}
\cline{1-8}
$\omega^2 \times 10^{3}$ & Fig. \ref{fig:1} urban & Fig. \ref{fig:1} suburban  & Fig. \ref{fig:2} $(a)$ & Fig. \ref{fig:2} $(b)$ & Fig. \ref{fig:3} & Fig. \ref{fig:4} $(a)$ & Fig. \ref{fig:4} $(b)$  \\
\hline
\hline
Lognormal & $2.254$ & $1.291$ & $2.941$ & $1.628$ & $0.043$ & $1.400$ & $1.796$\\
\hline
Gamma & $1.054$ & $1.181$ & $5.493$ & $1.099$ & $0.022$ & $1.730$ & $1.617$\\
\hline
Inverse Gamma & $10.583$&  $1.072$ & $1.520$ & $2.398$ & $0.021$ & $1.110$ & $2.691$\\
\hline
\makecell{Inverse Gamma \\ $m\in\mathbb{N}^+$} & $12.076$ & $1.075$ & $1.766$ & $2.448$ & $0.021$ & $1.111$ & $2.772$\\
\hline
Inverse Gaussian & $4.899$ & $1.094$ & $2.606$ & $1.712$ & $0.022$ & $1.387$ & $1.815$\\
\hline
\end{tabular}
\end{table*}

\begin{figure}[t]
	\centering
     \includegraphics[width=1\columnwidth]{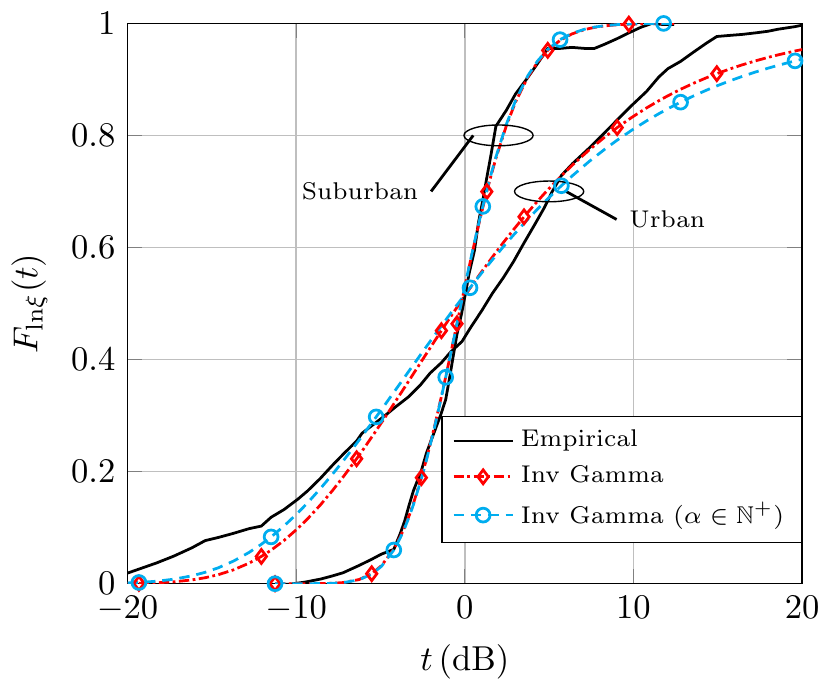}
         \caption{Impact of assuming $m \in \mathbb{N}^+$ in the inverse gamma distribution for data in \cite[Fig. 4]{Babiroli2017}, corresponding to urban and suburban scenarios at $169$ MHz. Parameters for urban scenario: inverse gamma ($m = 1.18$, $\Omega_i = 4.60$) and inverse gamma with integer $m$ ($m = 1$, $\Omega_i \to \infty$). Parameters for suburban scenario:  inverse gamma ($m = 9.82$, $\Omega_i = 1.05$) inverse gamma with integer $m$ ($m = 10$, $\Omega_i = 1.05$).}
      \label{fig:5}
\end{figure}

Aiming to cover a wide variety of propagation environments, we use empirical distributions obtained from data measurements corresponding to four different scenarios: urban and suburban scenarios for smart wireless metering systems at $169$ MHz \cite{Babiroli2017}, suburban at $910.25$ MHz \citep{Abdi1998}, and indoor scenarios at $26$ GHz \citep{Ai2017} and $45$ GHz \cite{Zhu2015} in order to also account for propagation effects at millimeter-wave frequency bands. 

The distributions' parameters in each scenario have been obtained by minimizing \eqref{eq:CramerTest} for each shadowing distribution in Sec. \ref{sec_IIA}. Since shadowing data are usually given in logarithmic scale (typically as deviations over the path loss in ${\rm dB}$), the fitting is not performed over the shadowing \ac{RV}, $\xi$, but over $\xi_{\textrm{ln}}={\rm ln}\,\xi$. Therefore, the corresponding change of variables is required in the theoretical \acp{CDF} in Definitions \ref{def:LN}-\ref{def:InvGamma}. Hence, we need to evaluate $F_\xi^M(\bm{\theta}_M; e^t)$, where $M$ denotes any of the considered distributions and $\bm{\theta}_M$ its set of parameters. Hence, the estimated set of parameters is obtained as
\begin{equation}
\bm{\widehat{\theta}}_M = \argmin\limits_{\bm{\theta}_M}\left\{\int_{-\infty}^\infty\left|\widehat{F}_{\xi_{\textrm{ln}}}(t) - F_\xi(\bm{\theta}_M;e^{t})\right|^2dt\right\}.
\end{equation}
Note that $\xi_{\textrm{ln}}$ is in natural logarithmic scale. That is, if the shadowing data is in dB scale, a rescaling factor must be applied as stated in [16], i.e., $t = 20 \,t_{\textrm{dB}}/ \textrm{ln}(10)$.

With this consideration in mind, the empirical and theoretical \acp{CDF} for each scenario are depicted in Figs. \ref{fig:1}-\ref{fig:4}, and the results for the Cramer-von Mises test are shown in Table \ref{Table1} (the distribution parameters for each case are detailed in the figure captions). Note that Table \ref{Table1} also provides the results for the \ac{IG} distribution when the parameter $m$ is restricted to be a positive integer, i.e., $m\in\mathbb{N}^+$. As we will later see, this consideration will have important benefits in the sequel, as it facilitates the statistical characterization of \ac{IG} based composite models; hence, it is important to quantify its practical impact in the fitting process to empirical data.

From Table \ref{Table1}, we observe that the widely used lognormal distribution is outperformed in all cases by either the gamma or the \ac{IG} models. More specifically, for severe shadowing --- corresponding to relatively large $\sigma$ for the lognormal distribution and small values of $k$ and $m$ for the gamma and \ac{IG} distributions, respectively --- the gamma distribution renders the best results. This is the case, for instance, of urban data in Fig. \ref{fig:1}. In turn, as the shadowing severity is reduced, the \ac{IG} arises as the best option (see, e.g., Fig. \ref{fig:2} case $(a)$ and Fig. \ref{fig:4} case $(b)$). In fact, it provides the most accurate fitting to the empirical \ac{CDF} in four of the cases under analysis, all of them corresponding to mild and moderate shadowing. 

Regarding now the case of the \ac{IG} distribution with integer $m$, the impact of this assumption seems negligible in those scenarios with medium and small shadowing variance (e.g., suburban data in Fig. \ref{fig:1}). This is a coherent result since mild shadowing implies large values of $m$, and then, rounding errors do not compromise much the fitting accuracy. In contrast, for small values of $m$ this impact is more notorious, as shown in Fig. \ref{fig:5}, where the \ac{CDF} of \ac{IG} model and \ac{IG} with integer $m$ are depicted for the data in Fig. \ref{fig:1}.

Based on these results, the use of the \ac{IG} distribution to model mild and moderate shadowing conditions arises as a reasonable choice, or at least as much as the gamma or lognormal distributions. In fact, it allows for an improved fitting accuracy even in the case of integer $m$. Most importantly, as we will now see, its mathematical tractability will lead to simpler expressions for the main statistics of \ac{IG}-based composite models than those resulting when considering other alternatives to model shadowing, such as lognormal or inverse Gaussian distributions.

\section{Physical model}
\label{sec:PhysicalModel}

Once the validation of the \ac{IG} distribution as shadowing model has been accomplished, we introduce in this section the physical model of composite fading distributions. Consider, therefore, a \ac{RV} $W$ characterizing the instantaneous received signal power in a multipath propagation scenario affected by both shadowing and fast fading. Then, $W$ can be expressed as
\begin{equation}
	W = \Omega \, \widehat{\xi} \widehat{X}
\end{equation}
where $\Omega$ is the mean signal power and $\widehat{\xi}$ and $\widehat{X}$ are independent \acp{RV} representing respectively the shadowing and the fast fading, with $\mathbb{E}[\widehat{\xi}] = \Omega_{\xi}$ and $\mathbb{E}[\widehat{X}] = \Omega_{X}$. As mentioned before, we consider in this paper the case in which $\widehat{\xi}$ is \ac{IG} distributed with shape parameter $m$ and \ac{PDF} $f_{\widehat{\xi}}^{\rm IG}(m, \Omega_\xi;t)$ given in \eqref{eq:InvGammaPDF}, whilst $\widehat{X}$ follows any arbitrary fading distribution. 

For the sake of simplicity, $W$ can be rewritten as
\begin{equation}
	\label{eq:W}
	W = \overline{W} \,\xi X
\end{equation} 
where $\overline{W} \triangleq \mathbb{E}[W] = \Omega\, \Omega_\xi\, \Omega_X$ and $\xi$ and $X$ are the normalized versions of $\widehat{\xi}$ and $\widehat{X}$. That is, $\xi$ is an \ac{IG} \ac{RV} with shape parameter $m$ and $\mathbb{E}[\xi] = 1$, i.e., its \ac{PDF} is given by $f_{\xi}^{\rm IG}(m, 1;t)$; and $X$ follows any fading distribution with $\mathbb{E}[X] = 1$. Hence, we work with normalized shadowing and fading models, whilst the impact of varying any of the powers is embedded in $\overline{W}$. Note that, as stated in the previous section, the value of $m$ is directly related to the severity of shadowing. Thus, smaller values of $m$ mean that the variance of the \ac{IG} distribution  --- equivalently, that of shadowing ---  is larger, while increasing $m$ renders less sparse values of $\xi$. This can also be observed from the set of parameters used in Figs. \ref{fig:1}-\ref{fig:4}.

\section{Statistical characterization of inverse gamma composite fading models}
\label{sec:Statistics}
\subsection{General case}
\label{sec:Statistics_General}

The current and the following subsections deal with the statistical characterization of composite models as defined in \eqref{eq:W}, aiming to provide a general methodology to obtain the \ac{PDF} and \ac{CDF} of $W$ in terms of the statistics of the underlying fast fading model, $X$, with independence of its distribution. Note that, although throughout the whole analysis we work with the statistics of the received signal power $W$, the \ac{PDF} and \ac{CDF} of the received signal amplitude $R$ can be straightforwardly derived from those of $W$ through a change of variables, rendering $f_R(r) = 2rf_W(r^2)$ and $F_R(r) = F_W(r^2)$, respectively.

We first consider the most general case in which shadowing, $\xi$, is \ac{IG} distributed with $m\in\mathbb{R}^+$ and $X$ follows an arbitrary fading distribution. Specifically, we will show that the first order statistics of $W$, namely the \ac{PDF} and \ac{CDF}, can be readily obtained from the \ac{GMGF} of $X$, which is defined below.   

\begin{definition}[GMGF]
	\label{def:GMGF}
	Let $X$ be a continuous non-negative \ac{RV} with \ac{PDF} $f_X(x)$ and consider $p\in\mathbb{R}^+$. Then, the \ac{GMGF} of $X$ is defined as
	\begin{equation}
		\label{eq:GMGF}
			\phi_X^{(p)}(s) \triangleq \mathbb{E}\left[X^p e^{Xs}\right] = \int_0^\infty x^p e^{xs} f_X(x) dx.
	\end{equation}
\end{definition}
Observe that, if $p\in\mathbb{N}^+$, then the \ac{GMGF} coincides with the $p^{th}$ order derivative of the \ac{MGF}, defined as $M_X(s) = \mathbb{E}[e^{sX}] = \phi_X^{(0)}(s)$. 

With Definition \ref{def:GMGF}, we now calculate the \ac{PDF} and \ac{CDF} of $W$ in the following lemmas. 

\begin{lemma}
	\label{lem:PDF}
	Let $W$ be a positive \ac{RV} characterizing the instantaneous received signal power as in \eqref{eq:W}. Then, for real $m>1$, its \ac{PDF} is given by
	\begin{equation}
		\label{eq:PDF}
		f_W(u) = \frac{\overline{W}^m(m-1)^m}{u^{m+1}\Gamma(m)}\phi_X^{(m)}\left(\frac{(1-m)\overline{W}}{u}\right).
	\end{equation}
\end{lemma}
\begin{IEEEproof}
	When conditioned on $\xi$, the \ac{PDF} of $W$ is 
	\begin{equation}
		f_W(u | \xi) = \frac{1}{\overline{W} \xi} f_X \left(\frac{u}{\overline{W} \xi}\right),
	\end{equation}
	with $f_X(\cdot)$ the \ac{PDF} of $X$. The unconditional \ac{PDF} is therefore obtained by averaging on $1/\xi$ as
	\begin{equation}
		\label{eq:PDFproof1}
		f_W(u) = \int_0^\infty \frac{1}{\overline{W}} t f_X \left(\frac{u \,t}{\overline{W}}\right) f_{1/\xi}(t) dt
	\end{equation}
	where, since $\xi$ is \ac{IG} distributed, then $1/\xi$ is gamma distributed with PDF $ f_{1/\xi}^{\rm G}(m, m/(m-1);t)$. Therefore, substituting in \eqref{eq:PDFproof1} and performing the change of variables $y = u t / \overline{W}$ lead to 
	\begin{equation}
		f_W(u) = \frac{\left[\overline{W}(m-1)\right]^m}{u^{m+1}\Gamma(m)}\int_0^\infty f_X(y) y^m e^{(1-m)\overline{W}y/u}dy.
	\end{equation}
	
	The above integral corresponds to the GMGF of $X$ evaluated at $s = (1-m)\overline{W}/u$, obtaining \eqref{eq:PDF} and completing the proof. 
\end{IEEEproof}

\begin{table*}[t]
\caption{GMGF of most commonly used fading distributions. ${}_1F_1(\cdot)$ denotes \mbox{the Kummer's confluent hypergeometric function \cite[eq. (13.1.2)]{Abra72}.}}
\label{Table2}
\centering
\setcellgapes{5pt}\makegapedcells
\begin{tabular}
{|c||c|}
\cline{1-2}
Fading model & $\phi_X^{(p)}(s) = \int_0^\infty x^p e^{xs} f_X(x) dx$, \quad$p\in\mathbb{R}^+$ \\
\hline
\hline
Rayleigh & $\phi_{\rm ray}^{(p)}(s) = \Gamma(p+1)\Omega_X^p(1-s\Omega_X)^{-(p+1)}$\\
\hline
Rician & $\phi_{\rm ric}^{(p)}(s) = \dfrac{\Gamma(p+1)\Omega_X^p (1+K)\exp(K)}{\left(1+K-s\,\Omega_X\right)^{p+1}}{}_1F_1\left(p+1; 1; \dfrac{K(1+K)}{1+K-s\,\Omega_X}\right)$\\
\hline
Nakagami-$\underline{m}$ & $\phi_{m}^{(p)}(s) = \dfrac{\Gamma(p+\underline{m})\Omega_X^p{\underline{m}}^{\underline{m}}}{\Gamma(\underline{m})(\underline{m}-s\Omega_X)^{p+\underline{m}}}$\\
\hline
Nakagami-$q$ (Hoyt) & $\phi_{ q}^{(p)}(s) = \dfrac{2^p q^{2p-1}\Gamma(p+1)\Omega_X^p(q^2+1)}{(q^2+1-2sq^2\Omega_X)^{p+1}}{}_2F_1\left(\dfrac{1}{2}, p+1; 1; \dfrac{1-q^4}{1+q^2-2sq^2\Omega_X}\right)$\\
\hline
$\kappa$-$\mu$ & $\phi_{\kappa\text{-}\mu}^{(p)}(s) = \dfrac{\Gamma(\mu+p)\Omega_X^p\mu^\mu (1+\kappa)^\mu \exp(\mu\kappa)}{\Gamma(\mu)\left(\mu(1+\kappa)-s\,\Omega_X\right)^{\mu+p}}{}_1F_1\left(\mu+p; \mu; \dfrac{\mu^2\kappa(1+\kappa)}{\mu(1+\kappa)-s\,\Omega_X}\right)$\\
\hline
$\eta$-$\mu$ (format 1) & $\phi_{\eta\text{-}\mu}^{(p)}(s) = \dfrac{\mu^{2\mu}\Gamma(p+2\mu)\Omega_X^p(\eta+1)^{2\mu}}{\eta^\mu\Gamma(2\mu)\left(\mu(\eta+1)/\eta-s\Omega_X\right)^{p+2\mu}}{}_2F_1\left(\mu, 2\mu+p; 2\mu; \dfrac{\mu(1-\eta^2)}{\mu(1+\eta)-s\eta\Omega_X}\right)$\\
\hline
\end{tabular}
\end{table*}

\begin{lemma}
	\label{lem:CDF}
	Let $W$ be a \ac{RV} characterizing the instantaneous received signal power as in \eqref{eq:W}. Then, for real $m>1$, its \ac{CDF} is given by 
	\begin{equation}
		\label{eq:CDF}
		F_W(u) = 1 - \sum_{n=0}^\infty \frac{\left[\overline{W}(m-1)\right]^{m+n}}{u^{m+n}\Gamma (m+n+1)}\phi_X^{(m+n)}\left(\frac{(1-m)}{u \overline{W}^{-1}}\right).
	\end{equation}
\end{lemma}
\begin{IEEEproof}
	Similarly to the PDF, the CDF of $W$ can be calculated as 
	\begin{equation}
		\label{eq:CDFproof2}
		F_W(u) = \int_0^\infty F_X\left(\frac{u\,t}{\overline{W}}\right)f_{1/\xi}(t) dt.
	\end{equation}
	Performing the change of variables $y = ut/\overline{W}$ and integrating by parts we obtain 
	\begin{equation}
		\label{eq:CDFproof1}
		F_W(u) = 1-\int_0^\infty F_{1/\xi}\left(\frac{\overline{W}}{u}y\right)f_X(y)dy,
	\end{equation}
	where $F_{1/\xi}(\cdot)$ is the CDF of $1/\xi$, which is gamma distributed with shape parameter $m$ and $\Omega = m/(m-1)$. Therefore, using \eqref{eq:GammaCDF} and \cite[eqs. (6.5.4) and (6.5.24)]{Abra72}, we can rewrite \eqref{eq:CDFproof1} as
	\begin{align}
		F_W(u) =& 1 - \sum_{n=0}^\infty \frac{\left[\overline{W}(m-1)\right]^{m+n}}{u^{m+n}\Gamma (m+n+1)}\notag \\ 
		&\times  \int_0^\infty y^{m+n}e^{(1-m)\overline{W}y/u}f_X(y)dy,
	\end{align}
	where the integral corresponds to the GMGF of $X$ given in \eqref{eq:GMGF}, yielding \eqref{eq:CDF}.
\end{IEEEproof}

Lemmas \ref{lem:PDF} and \ref{lem:CDF} introduce general expressions for the \ac{PDF} and \ac{CDF} of \ac{IG}-based composite fading distributions in terms of the \ac{GMGF} of the underlying fading model. These results, which are new in the literature to the best of the authors' knowledge, provide an unified methodology to characterize composite models. Hence, once the \ac{GMGF} of the fading distribution is obtained, the statistical analysis of the composite version is straightforward. Indeed, the analytical tractability of \eqref{eq:PDF} and \eqref{eq:CDF} will strongly depend on the ability to calculate the \ac{GMGF} of $X$. 

Despite not being as popular as the \ac{MGF}, the \ac{GMGF} of many fading models can be obtained in closed-form for arbitrary $p$. This is exemplified, for instance, by the very general $\kappa$-$\mu$ shadowed distribution \cite{Paris2014,Cotton2015}, which includes most popular fading distributions as particular cases, and whose \ac{GMGF} is provided in the following lemma.

\begin{lemma}
	\label{lem:GMGF_KMS}
	Consider a \ac{RV}, $X$, following a $\kappa$-$\mu$ shadowed distribution with parameters $\kappa$, $\mu$, $\underline{m}$\footnote{Note that the parameter $\underline{m}$, inherent to the $\kappa$-$\mu$ shadowed distribution, is underlined in order not to be confused with that of the IG distribution.} and $\mathbb{E}[X]=\Omega_X$. Then, its \ac{GMGF} is given by 
\begin{align}
    \phi^{(p)}_{\kappa\text{-}\mu \mathcal{S}}(s) &= \frac{\Gamma(\mu+p)\underline{m}^{\underline{m}}\Omega_X^p\mu^\mu (1+\kappa)^\mu}{\Gamma(\mu)(\mu\kappa+\underline{m})^{\underline{m}}\left(\mu(1+\kappa)-s\,\Omega_X\right)^{\mu+p}}\notag \\
    &\times{}_2F_1\left(\underline{m},\mu+p; \mu; \frac{\mu^2\kappa(1+\kappa)(\mu\kappa+\underline{m})^{-1}}{\mu(1+\kappa)-s\,\Omega_X}\right), \label{eq:GMGF_KMS} 
\end{align}
where ${}_2F_1(\cdot)$ is the Gauss' hypergeometric function \cite[eq. (15.1.1)]{Abra72}.
\end{lemma}
\begin{IEEEproof}
	Introducing the \ac{PDF} of the $\kappa$-$\mu$ shadowed distribution given by \cite[eq. (4)]{Paris2014} in \eqref{eq:GMGF}, \eqref{eq:GMGF_KMS} is straightforwardly obtained by using \cite[eq. (7.621 4)]{Gradshteyn07} and performing some algebraic manipulations.  
\end{IEEEproof}

From \eqref{eq:GMGF_KMS}, the \ac{GMGF} of most commonly used fading distributions are obtained just by applying the relationships between the $\kappa$-$\mu$ shadowed model and the classical and generalized distributions derived from it. These connections are given, e.g., in \cite[Table 1]{Moreno2016}, and the resulting \acp{GMGF} for the distinct particular cases are summarized in Table \ref{Table2}, at the top of this page. 

As shown in Table \ref{Table2} and Lemma \ref{lem:GMGF_KMS}, the \acp{GMGF} of most widely used distributions are readily calculated, and therefore the characterization of \ac{IG} composite version of a very large number of models is direct. Moreover, in the most general case where the \ac{GMGF} of the considered fading distribution is unknown or has an intractable form, the integral in \eqref{eq:GMGF} can be computed numerically, as it is generally well-behaved since the exponential term should ensure the convergence.

\subsection{The case of integer $m$}
\label{sec:Statistics_Integer}

Results in the previous subsection are valid for any arbitrary $m\in\mathbb{R}^+$, i.e., the shape parameter of the \ac{IG} \ac{RV} characterizing the shadowing can take any positive real value. However, the tractability of both the gamma and \ac{IG} distributions considerably improves when assuming that the shape parameter is a positive integer number. More specifically, the \ac{CDF} of both distributions is given in terms of the incomplete gamma function \eqref{eq:GammaCDF}, \eqref{eq:InvGammaCDF}, which admits a simple closed-form representation in such case. Hence, assuming $a\in\mathbb{N}^+$, then  
\begin{equation}
	\label{eq:GamInc_Integer}
	\frac{\gamma(a,z)}{\Gamma(a)} = 1-\frac{\Gamma(a,z)}{\Gamma(a)} = 1 - e^{-z}\sum_{k=0}^{a-1} \frac{z^k}{k!}, 
\end{equation}
which is easily proved by using \cite[eq. (3.351 1)]{Gradshteyn07}. 

The improved mathematical tractability of the \ac{IG} distribution under the aforementioned assumption directly translates into a simplified expression for the \ac{CDF} of the instantaneous received signal power. Thus, the \ac{CDF} of $W$ is no longer given by a infinite series but instead by a finite sum of evaluations of the \ac{GMGF} of the underlying fading model $X$, as stated in the following corollary:

\begin{corollary}
	\label{cor:CDF}
	Let $W$ be a RV characterizing the instantaneous received signal power in \eqref{eq:W}, and assume $m$ is a positive integer, i.e, $m\in\mathbb{N}^+$. Then, the CDF of $W$ is expressed as  
	\begin{equation}
		\label{eq:CDFinteger}
		F_W(u) = \sum_{n=0}^{m-1}\frac{(m-1)^n\overline{W}^n}{u^n \Gamma(n+1)}\phi_X^{(n)}\left(\frac{(1-m)\overline{W}}{u}\right).
	\end{equation}
\end{corollary}
\begin{IEEEproof}
	Since $1/\xi$ is gamma distributed with shape parameter $m$ and $\mathbb{E}[1/\xi] = m/(m-1)$, we can particularize its \ac{CDF} for integer $m$ by introducing \eqref{eq:GamInc_Integer} in \eqref{eq:GammaCDF}, leading to
	\begin{equation}
		\label{eq:CDFxi_integer}
		F_{1/\xi}(t) = 1 - \sum_{k=0}^{m-1}\frac{(m-1)^k t^k}{\Gamma(k+1)}e^{(1-m)t}.
	\end{equation}
	Substituting \eqref{eq:CDFxi_integer} in \eqref{eq:CDFproof1} and following the same steps as in Lemma \ref{lem:CDF}, the proof is completed.
\end{IEEEproof}

Note that, as shown in Section \ref{sec:Fitting}, assuming $m\in\mathbb{N}^+$ has a negligible effect in practice from a goodness-of-fit perspective --- unless the shadowing variance is large --- so we can achieve an improved tractability without compromising the accuracy of the model.

Assuming integer $m$ has also additional benefits. For instance, the \ac{GMGF} of more sophisticated fading models such as Beckmann or \ac{TWDP}, although not available for the general case, have already been obtained for $m\in\mathbb{N}^+$ --- equivalently, $p\in\mathbb{N}^+$ in \eqref{eq:GMGF} --- in closed-form \cite{Pena2017, Pena2018}. 

Strikingly, even though neither the Beckmann nor the \ac{TWDP} distributions admit closed-form expressions for their \ac{PDF} or \ac{CDF}, their respective composite models do. Hence, somehow counterintuitively, the \ac{IG} distribution not only renders more general models, but at the same time their mathematical complexity is even relaxed.


\subsection{The case with a fading distribution as a mixture of gammas}
\label{sec:Statistics_Mixture}

Some general fading models, albeit being very versatile, may pose a challenge from an analytical point of view since their statistics (mainly \ac{PDF} and \ac{CDF}) are given by intricate expressions; i.e, usually involving complicated special functions, or even in integral form as in the \ac{TWDP} case \cite{Durgin2002}. A classical solution to this issue is expressing the target distribution as a mixture of more tractable distributions.

Approximations based on mixtures are a well-known technique that has been widely used in approximation theory and applied statistics to characterize intricate \acp{RV} using different baseline distributions \cite{Everitt1981, Wiper2001, Sorenson1971}. In channel modeling, the preferred choice is the gamma distribution, due to its mathematical tractability and the simplicity of the resulting mixture statistics, which allow for the calculation of relevant performance metrics in wireless communication systems \cite{Atapattu2011}. In this case, the \ac{PDF} of the considered fading model is aimed to be expressed as \cite[eq. (1)]{Atapattu2011}:
\begin{equation}
	\label{eq:PDFMixGamma}
	f_X(u) = \sum_{i=1}^N w_i f_i^{\rm G}(k_i,\Omega_i; u)
\end{equation}
where $f_i^{\rm G}(\cdot)$ is the gamma PDF in \eqref{eq:GammaPDF}, the mixture coefficients $w_i$ for $i=1,\dots,N$ are constants that satisfy $\sum_{i=1}^{N}w_i=1$, and $k_i$ and $\Omega_i$ are the parameters of the $i$-th gamma distribution. 

In some cases, this mixture form naturally arises by inspection. This is the case, for instance, of the $\kappa$-$\mu$ shadowed fading model, whose \ac{PDF} can be written as 
\begin{equation}
	\label{eq:KMSPDF_mixture}
	f_{\kappa\text{-}\mu \mathcal{S}}(u) = \sum_{i=0}^\infty w_i^ \mathcal{S} f_i^{\rm G}\left(\mu+i,\frac{\Omega_X(\mu+i)}{\mu(\kappa+1)}; u\right)
\end{equation}
with $\Omega_X = \mathbb{E}[X]$ and
\begin{equation}
	w_i^ \mathcal{S} = \frac{\Gamma(\underline{m}+i)}{\Gamma(\underline{m}) \Gamma(i+1)} \frac{(\mu\kappa)^i \underline{m}^{\underline{m}}}{(\mu\kappa + \underline{m})^{\underline{m}+i}}.
\end{equation}

The above expression is easily proved by inspection after introducing \cite[eq. (13.1.2)]{Abra72} in the \ac{PDF} of the $\kappa$-$\mu$ shadowed distribution given by \cite[eq. (4)]{Paris2014} and after performing some algebraic manipulations. Note that, to the best of our knowledge, \eqref{eq:KMSPDF_mixture} is new in the literature. Notably, when both $\underline{m}$ and $\mu$ are positive integers, then \eqref{eq:KMSPDF_mixture} reduces to a finite mixture as proved in \cite{Lopez2017}. Logically, as with the \ac{GMGF}, from \eqref{eq:KMSPDF_mixture} and \cite{Lopez2017}, any fading distribution regarded as a particularization of the $\kappa$-$\mu$ model can also be expressed as a mixture of gammas. Another interesting case is that of the \ac{TWDP} model, whose \ac{PDF} is given in integral form but also admits an exact mixture representation as provided in \cite{Ermolova2016}, considerably improving its tractability. 

In all instances, whenever the \ac{PDF} of the fading model can be expressed as a mixture of gamma distributions, the \ac{PDF} of the \ac{IG}-based composite fading model is directly obtained as a mixture of $\mathcal{F}$ distributions as stated in the following lemma:

\begin{lemma}
	\label{lem:Mixture}
	Let $X$ be a \ac{RV} characterizing the fast fading with \ac{PDF} as in \eqref{eq:PDFMixGamma}. Then, the \ac{PDF} of \ac{IG} composite model is given by 
	\begin{equation}
		f_W(u) = \sum_{i=1}^N w_i f_i^\mathcal{F}(m, k_i, \Omega_i; u),
	\end{equation}
	where $f_i^{\mathcal{F}}(\cdot)$ is the \ac{PDF} of the $\mathcal{F}$ distribution \cite[eq. (6)]{Yoo2019}:
	\begin{equation}
		\label{eq:PDF_F}
		f^{\mathcal{F}}(m, k, \Omega; t) = \frac{(m-1)^m k^k}{B(m,k)}\frac{t^{k-1}\Omega^m}{\left((m-1)\Omega + kt\right)^{m+k}}
	\end{equation}
	with $B(\cdot,\cdot)$ the beta function \cite[eq. (6.2.2)]{Abra72}.
\end{lemma}
\begin{IEEEproof}
	From \eqref{eq:PDFMixGamma}, the \ac{PDF} of $W$ is calculated as
	\begin{equation}
		f_W(u) = \sum_{i=1}^N w_i \int_0^\infty t f_i^{\rm G}(k_i, \Omega_i; ut)f_{1/\xi}(t) dt,
	\end{equation}
	with $f_i^{\rm G}(\cdot)$ as in \eqref{eq:GammaPDF} and $f_{1/\xi}(t) = f_{1/\xi}^{\rm G}(m, m/(m-1);t)$. Using \cite[eq. (3.381 4)]{Gradshteyn07} and performing some algebraic manipulations, the proof is completed.
\end{IEEEproof}

The above lemma provides a remarkable result, since whenever the underlying fading model $X$ can be expressed as in \eqref{eq:PDFMixGamma}, then the statistical characterization of the composite model is straightforward, as we can leverage all the existing results given for the reasonably simple $\mathcal{F}$ distribution.  

\section{Outage Analysis in IG based composite models}
\label{sec:OutageAnalysis}

In the previous section, we have introduced a complete mathematical framework to characterize \emph{any} composite fading model based on \ac{IG} shadowing. Now, we aim to show how these previous results translate into the analysis of one of the key metrics in wireless communications: the outage probability. 

From \eqref{eq:W}, the instantaneous \ac{SNR} at the receiver is expressed as
\begin{equation}
	\label{eq:SNR}
	\gamma = \overline{\gamma} W/\overline{W}, 
\end{equation}
where, by definition, $\overline{\gamma} = \mathbb{E}[\gamma]$. Then, considering $\gamma_{\textrm{th}}$ as the minimum \ac{SNR} required for a reliable communication, the outage probability is given by \cite[eq. (6.46)]{Goldsmith05}
\begin{equation}
	P_{\textrm{out}}(\gamma_{\textrm{th}}) \triangleq P(\gamma < \gamma_{\textrm{th}}) = F_\gamma(\gamma_{\textrm{th}}),
\end{equation}
which can be straightforwardly obtained from the \ac{CDF} of $W$ as $F_\gamma(\gamma_{\textrm{th}}) = F_W(\overline{W}\gamma_{\textrm{th}} / \overline{\gamma})$. Therefore, expressions for the outage probability over arbitrary IG-based composite models are available, e.g., in \eqref{eq:CDF} for the general case and in \eqref{eq:CDFinteger} for the integer $m$ case. 

Although the previous result provides a complete characterization of the outage probability, a more insightful analysis can be carried out by considering that $\overline{\gamma}$ is large enough. Note that this scenario is of interest in, e.g., \ac{URLLC} \cite{Eggers2019}. Then, assuming $\overline{\gamma}\rightarrow\infty$, the outage probability is determined by the left tail of the distribution, which for the majority of fast fading models follows a power law given by \cite{Wang2003}
\begin{equation}
	\label{eq:FXasymp}
	F_{\widehat{X}} (x\,|\,x\rightarrow 0^+) = \frac{\alpha}{\beta+1} \left(\frac{x}{\Omega_X}\right)^{\beta+1},
\end{equation}
with $\alpha$ and $\beta$ are non-negative real constants depending on the fading distribution. The values of $\alpha$ and $\beta$ for a wide variety of both classical and generalized models can be found in \cite{Zhu2018, Eggers2019}. With this in mind, the asymptotic outage probability for IG-based composite models is provided in the following lemma.
\begin{lemma}
	Consider $\gamma$ as in \eqref{eq:SNR} with $W$ as in \eqref{eq:W}, and assume that the \ac{CDF} of the underlying fast fading model, $\widehat{X}$, admits the formulation in \eqref{eq:FXasymp}. Then, for $m>1$, the asymptotic outage probability over $\gamma$ is given by
	\begin{equation}
		\label{eq:OutageAsymp}
			P_{\textrm{out}}(\gamma_{\textrm{th}} | \overline{\gamma}\rightarrow \infty) = \frac{\Gamma(\beta + m + 1)}{\Gamma(m)(m-1)^{\beta+1}} \frac{\alpha}{\beta+1} \left(\frac{\gamma_{\textrm{th}}}{\overline{\gamma}}\right)^{\beta+1}.
	\end{equation}
\end{lemma}
\begin{IEEEproof}
	Introducing \eqref{eq:FXasymp} in \eqref{eq:CDFproof2}, using \cite[eq. (3.381 4)]{Gradshteyn07} and performing the corresponding change of variables, the proof is completed. 
\end{IEEEproof}

Remarkably, the composite model exhibits the same diversity order ($D=\beta + 1$) than the original fading distribution, and the impact of the shadowing in the large \ac{SNR} regime reduces to an scaling factor. Note that this effect is also observed in composite models based on the lognormal distribution, as proved in \cite{Zhu2018}. Hence, approximating the lognormal shadowing by the \ac{IG}-based one remains unaltered the asymptotic properties of the outage probability. 


\section{Application: a composite IG/TWDP fading model}
\label{sec:TWDPIG}
As a by-product of the theoretical formulation introduced in the previous sections, a number of composite fading models arises. For instance, a composite IG/$\kappa$-$\mu$ shadowed fading model (and all special cases included therein) is directly obtained for arbitrary values of the shape parameter $m$. We now aim to provide an additional example to illustrate the usefulness of our approach to build composite fading models based on IG shadowing. 

For this purpose, the case of the TWDP fading model is considered, which assumes the presence of two dominant specular components and accurately fits field measurements in a variety of propagation scenarios \cite{Durgin2002}.

\begin{figure*}[!t]
\setcounter{equation}{36}
\normalsize
\begin{align}
		\phi_{W_T}^{(p)}(K,\Delta,\Omega_T; s \,|\, p\in\mathbb{N}^+) =& \frac{\overline{W}^p}{\pi} \sum_{q=0}^p \frac{p!}{q!}\binom{p}{q} \frac{K^q(K+1)^{q+1}}{(K+1-\overline{W}s)^{p+q+1}} \exp\left(\frac{K\overline{W}s}{K+1-\overline{W}s}\right)\sum_{j=0}^q \binom{q}{j}\frac{\Delta^j}{2} \notag \\
		&\times \left[B\left(\frac{j+1}{2},\frac{1}{2}\right){}_1F_2\left(\frac{j+1}{2}; \frac{1}{2}, \frac{j+2}{2}; \frac{1}{4}\left(\frac{K\Delta\overline{W}s}{K+1-\overline{W}s}\right)^2\right)\left(1+(-1)^j\right)\right. \notag \\
		&\left. + \frac{K\Delta\overline{W}s}{K+1-\overline{W}s}B\left(\frac{j+2}{2}, \frac{1}{2}\right){}_1F_2\left(\frac{j+2}{2};\frac{3}{2}, \frac{j+3}{2};\frac{1}{4}\left(\frac{K\Delta\overline{W}s}{K+1+\overline{W}s}\right)^2\right)(1-(-1)^j)\right]. \label{eq:GMGF_TWDP}
\end{align}
\hrulefill
\vspace*{4pt}
\end{figure*}

\subsection{TWDP fading distribution}
According to the \ac{TWDP} fading model, the received signal power $W_{T}$ is described as
\setcounter{equation}{33}
\begin{equation}
	\label{TWDP_physical}
	W_T = \left|V_1 e^{j\varphi_1}+ V_2 e^{j\varphi_2} + Z\right|^2,
\end{equation}
where $V_1\in\mathbb{R}^+$ and $V_2\in\mathbb{R}^+$ are constants representing the amplitude of each specular component, $\varphi_i$ and $\varphi_2$ are \acp{RV} following an uniform distribution, i.e., $\varphi_{1,2}\sim\mathcal{U}[0,2\pi)$, and $Z$ is a complex Gaussian \ac{RV} such that $Z\sim\mathcal{CN}(0,2\sigma^2)$. It is assumed that all the involved \acp{RV} are statistically independent. The model is completely described by the parameters 
\begin{align}
	K &= \frac{V_1^2+V_2^2}{2\sigma^2}, & \Delta &= \frac{2V_1V_2}{V_1^2+V_2^2}.
\end{align}

Analogously to the Rician parameter, $K$ represents the ratio between the powers of the specular and the diffuse components, whilst $\Delta$ accounts for the difference between the power of each specular component. Thus, $\Delta = 1$ implies that $V_1 = V_2$ while $\Delta = 0$ makes one of the direct rays to vanish in \eqref{TWDP_physical}, obtaining the Rician distribution.

Although flexible, the \ac{TWDP} model is challenging from a analytical point of view since its \ac{PDF} is given in integral form involving the product of several Bessel's functions \cite[eq. (7)]{Durgin2002}. An alternative formulation is provided in \cite[eq. (16)]{Rao2015}, where the \ac{PDF} is expressed as\footnote{Note that there is a typo in \cite[eq. (16)]{Rao2015}, which has been corrected here.}
\begin{align}
		f_{W_T}(K,&\Delta,\Omega_T;u) = \frac{1+K}{2\pi\Omega_T} \exp\left(-\frac{(1+K)u}{\Omega_T}-K\right) \notag \\
		&\times \int_0^{2\pi} e^{-K\Delta\cos \alpha} I_0\left(2\sqrt{\frac{ u (1+\Delta\cos \alpha)}{K^{-1}(1+K)^{-1}\Omega_T}}\right)d\alpha,
		\label{eq:TWDP_pdf}
\end{align}
with $\Omega_T = \mathbb{E}[W_T] = V_1^2 + V_2^2 + 2\sigma^2$. Note that, despite being also in integral form, it is much simpler to compute \eqref{eq:TWDP_pdf} than \cite[eq. (7)]{Durgin2002}. Other formulations are also given in \cite[eq. (17)]{Durgin2002} and \cite{Saberali2013}, where approximated series expansions for $f_{W_T}(u)$ are derived, and in \cite[eq. (6)]{Ermolova2016}, where the \ac{PDF} is expressed as a mixture of gamma densities as
\setcounter{equation}{37}
\begin{equation}
	\label{eq:TWDP_mixture}
	f_{W_T}(K,\Delta, \Omega_T; u) = e^{-K}\sum_{j=0}^\infty w^T_j f_j^{\rm G}\left(j+1, \frac{(j+1)\Omega_T}{K+1}; u\right),
\end{equation} 
with 
\begin{equation}
	\label{eq:w_twdp}
	w^T_j = \frac{K^j}{j!}\sum_{i=0}^j \binom{j}{i}\left(\frac{\Delta}{2}\right)^i \sum_{l=0}^i \binom{i}{l}I_{2l-i}(-K\Delta),
\end{equation}
where $I_\nu(\cdot)$ is the modified Bessel's function \cite[eq. (8431)]{Gradshteyn07}.

\subsection{Inverse gamma/TWDP composite fading model}
With the \ac{TWDP} distribution as baseline model, the composite fading model \ac{IG}/\ac{TWDP} is built according to \eqref{eq:W} as
\begin{equation}
	\label{eq:IGTWDP_physical}
	W = \overline{W}\xi W_T,
\end{equation}
where $\xi$ follows an inverse gamma distribution with shape parameter $m$ and $\mathbb{E}[\xi] = 1$ and $\mathbb{E}[W_T] = 1$. To illustrate the usefulness of our proposed approach to characterize \ac{IG}-based composite models, we will provide expressions for both the \ac{PDF} and the \ac{CDF} in all the considered cases: i) the general case ($m\in\mathbb{R}^+$), ii) the particular case in which $m\in\mathbb{N}^+$, and iii) the mixture case. 

\subsubsection{The case $m\in\mathbb{R}^+$}

For arbitrary $m\in\mathbb{R}^+$ (equivalently, $p\in\mathbb{R}^+$), no closed-form expression is available in the literature for the \ac{GMGF} of the \ac{TWDP} fading model, having to resort to the integral definition in \eqref{eq:GMGF}. Therefore, substituting the \ac{GMGF} definition in \eqref{eq:PDF} and \eqref{eq:CDF} we obtain expressions for the \ac{PDF} and \ac{CDF} of the \ac{IG}/\ac{TWDP} in terms of a double integral. Interchanging the order of integration and using \cite[eqs. (6.643 2) and (9.220 2)]{Gradshteyn07} we finally get
\begin{align}
	f_W(u) =& \frac{e^{-K}m\overline{W}^m(1+K)(m-1)^m}{2\pi\left[(m-1)\overline{W}+(1+K)u\right]^{m+1}}\int_{0}^{2\pi}e^{-K\Delta\cos(\alpha)} \notag \\
	&\times {}_1F_1\left(m+1;1;\frac{uK(1+\Delta\cos(\alpha))}{u+\frac{(m-1)\overline{W}}{1+K}}\right)\,d\alpha, \label{eq:PDFTWDP_general}
\end{align}
\begin{align}
	F_W(u) =& 1-\sum_{n=0}^\infty\frac{[\overline{W}(m-1)]^{m+n}e^{-K}(1+K)u}{2\pi[(m-1)\overline{W}+(1+K)u]^{m+n+1}} \notag \\
	&\times \int_0^{2\pi}{}_1F_1\left(m+n+1;1;\frac{uK(1+\Delta\cos(\alpha))}{u+\frac{(m-1)\overline{W}}{1+K}}\right) \notag \\
	&\times e^{-K\Delta\cos(\alpha)} \,d\alpha, \label{eq:CDFTWDP_general}
\end{align}
where we reduced the twofold integral to a simple one in a closed interval. Evaluating \eqref{eq:CDFTWDP_general} may be seem challenging at a first glance from a computational point of view, as it requires to compute several integrals numerically. However, they are well-behaved and therefore no numerical issues arise in standard calculation software such as \MATLAB or \MATHE.

\subsubsection{The case $m\in\mathbb{N}^+$}

As stated before, considering $m$ to be a positive integer renders considerable benefits from an analytical point of view. In this case, the \ac{GMGF} of the \ac{TWDP} distribution admits a closed-form expression given in \cite{Pena2018} and reproduced in \eqref{eq:GMGF_TWDP}, at the top of this page, for reader's convenience. In \eqref{eq:GMGF_TWDP}, ${}_1F_2(\cdot)$ is a generalized confluent hypergeometric function \cite[p. 19]{Srivastava1985}. Therefore, the \ac{PDF} and \ac{CDF} of $W$ in \eqref{eq:IGTWDP_physical} are straightforwardly calculated by introducing \eqref{eq:GMGF_TWDP} in \eqref{eq:PDF} and \eqref{eq:CDFinteger}, respectively, obtaining
\begin{align}
	f_W(u) =& \frac{\overline{W}^m(m-1)^m}{u^{m+1}\Gamma(m)}\phi_{W_T}^{(m)}\left(K,\Delta,1; \frac{(1-m)}{u\overline{W}^{-1}}\right),\label{eq:PDFTWDP_integer} \\[1ex]
F_W(u) =& \sum_{n=0}^{m-1}\frac{(m-1)^n\overline{W}^n}{u^n \Gamma(n+1)}\phi_{W_T}^{(n)}\left(K,\Delta,1; \frac{(1-m)}{u\overline{W}^{-1}}\right).
\label{eq:CDFTWDP_integer}
\end{align}

Note that the above expressions are easier to compute than \eqref{eq:PDFTWDP_general} and \eqref{eq:CDFTWDP_general}, since not only the \ac{GMGF} admits a closed-form expression but also the \ac{CDF} is no longer expressed as an infinite sum. 

\subsubsection{The mixture case}

The last option is using the mixture approach in \eqref{eq:TWDP_mixture}, in which the \ac{PDF} of the \ac{TWDP} is expressed as a mixture of gamma \acp{PDF}. Once again, the analysis of the composite version is straightforward by using the results presented in this paper. Thus, the \ac{PDF} of the \ac{IG}/\ac{TWDP} model is readily derived from Lemma \ref{lem:Mixture} as
\begin{align}
	\label{eq:IGTWDP_mixturePDF}
	f_W(u) = e^{-K}\sum_{j=0}^\infty w_j^T f_j^{\mathcal{F}}\left(m,j+1, \frac{(j+1)\overline{W}}{K+1}; u\right),
\end{align}
where $f^{\mathcal{F}}(\cdot)$ is the \ac{PDF} of the $\mathcal{F}$ distribution in \eqref{eq:PDF_F}. Similarly, the \ac{CDF} is easily calculated by integrating \eqref{eq:IGTWDP_mixturePDF}, obtaining
\begin{equation}
\label{eq:IGTWDP_mixtureCDF}
	F_W(u) = e^{-K}\sum_{j=0}^\infty w_j^T F_j^{\mathcal{F}}\left(m,j+1, \frac{(j+1)\overline{W}}{K+1}; u\right),
\end{equation}
with $F^{\mathcal{F}}(\cdot)$ being the $\mathcal{F}$ \ac{CDF} \cite[eq. (12)]{Yoo2019}:
\begin{align}
	F^{\mathcal{F}}(m,k,\Omega;t) &= \frac{k^{k-1}t^k}{B(m,k)(m-1)^k\Omega^k} \notag \\
	&\times{}_2F_1\left(k,k+m;k+1;\frac{-kt}{(m-1)\Omega}\right). 
\end{align}

\subsubsection{Outage probability}

Considering that the \ac{SNR} is given by \eqref{eq:SNR} with $W$ as in \eqref{eq:IGTWDP_physical}, the outage probability is therefore given by $P_{\textrm{out}}(\gamma_{\textrm{th}}) = F_W(\overline{W}\gamma_{\textrm{th}} / \overline{\gamma})$. If we take the assumption $\overline{\gamma}\rightarrow\infty$, then \eqref{eq:OutageAsymp} can be applied. In the \ac{TWDP} case, the power law parameters are given by \cite[eq. (33)]{Garcia2019}
\begin{align}
\alpha &= (1+K)e^{-K}I_0(K\Delta), & \beta &= 0,
\end{align}
obtaining 
\begin{equation}
	\label{eq:IGTWDP_asym}
	P_{\textrm{out}}(\gamma_{\textrm{th}} | \overline{\gamma}\rightarrow \infty) = \frac{m(1+K)I_0(K\Delta)}{(m-1)e^{K}} \frac{\gamma_{\textrm{th}}}{\overline{\gamma}}.
\end{equation}

\subsection{Numerical results}
\begin{figure}[t]
	\centering
     \includegraphics[width=\columnwidth]{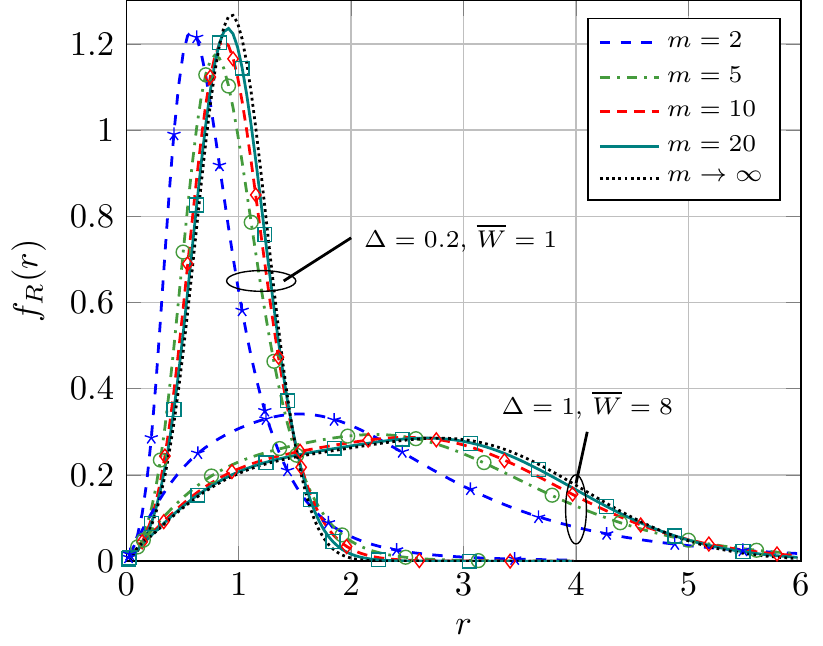}
         \caption{\ac{PDF} of the received signal amplitude under \ac{IG}/\ac{TWDP} fading for $K=4$ and different values of $m$, $\Delta$ and $\overline{W}$. Solid lines correspond to theoretical calculations using the general expression in \eqref{eq:PDFTWDP_general}, whilst markers correspond to \ac{MC} simulations.}
      \label{fig:6}
\end{figure}

\begin{figure}[t]
	\centering
     \includegraphics[width=\columnwidth]{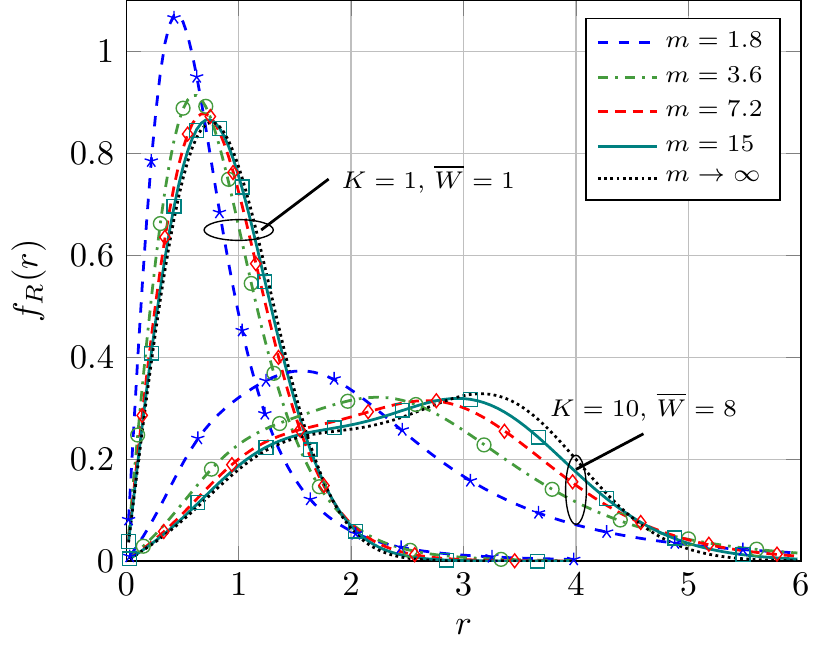}
         \caption{\ac{PDF} of the received signal amplitude under \ac{IG}/\ac{TWDP} fading for $\Delta=0.9$ and different values of $m$, $K$ and $\overline{W}$. Solid lines correspond to theoretical calculations using the mixture representation in \eqref{eq:IGTWDP_mixturePDF} with $N=15$ terms computed for $\overline{W}=1$ and $N=30$ for $\overline{W}=8$, whilst markers correspond to \ac{MC} simulations.}
      \label{fig:7}
\end{figure}

\begin{figure}[t]
	\centering
     \includegraphics[width=\columnwidth]{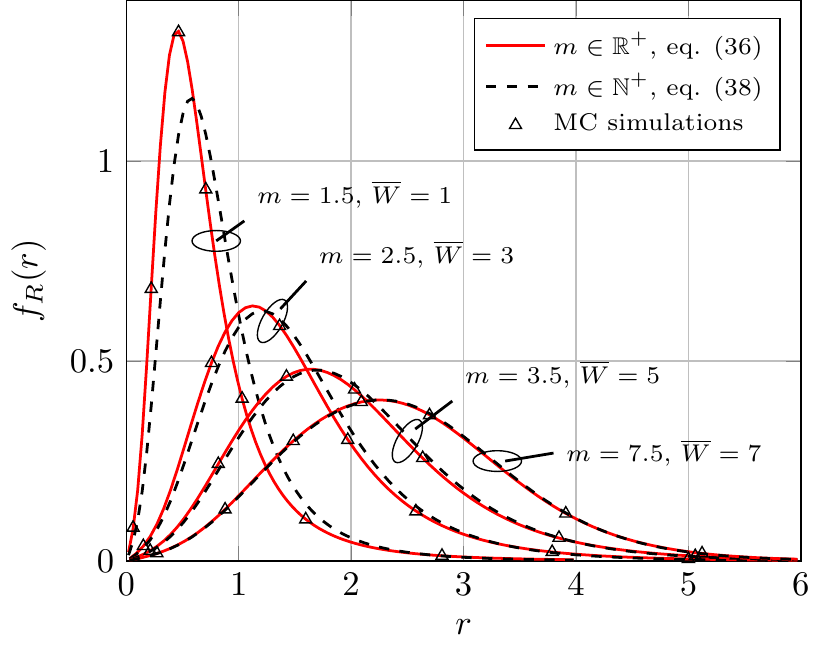}
         \caption{Impact of considering integer $m$ in the \ac{PDF} of the received signal amplitude under \ac{IG}/\ac{TWDP} fading for $K=7$ and $\Delta=0.7$. The value of $m$ in the integer cases have been obtained by rounding it. For better clarity, different values of $\overline{W}$ have been used in all cases.}
      \label{fig:8}
\end{figure}

\begin{figure}[t]
	\centering
     \includegraphics[width=\columnwidth]{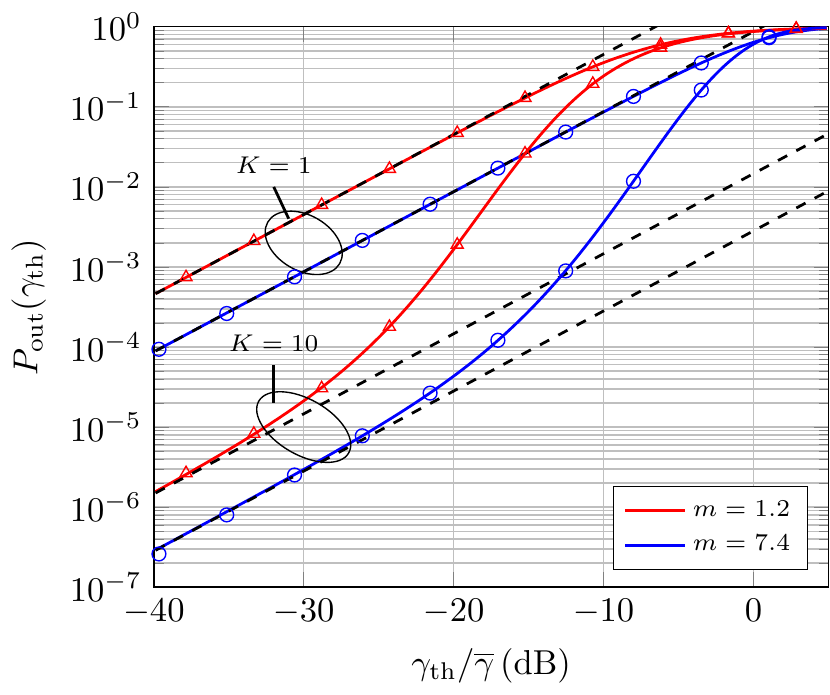}
         \caption{Outage probability under \ac{IG}/\ac{TWDP} fading for $\Delta = 0.3$ and different values of $m$, $K$. Solid lines correspond to theoretical calculations using the mixture representation in \eqref{eq:IGTWDP_mixtureCDF} with $N=20$ terms computed, dashed lines correspond to the asymptotic expression in \eqref{eq:IGTWDP_asym} and markers correspond to \ac{MC} simulations.}
      \label{fig:9}
\end{figure}

Aiming to both visualize the impact of $m$ in the distribution of $W$ and to check the validity of the derived results, we show in Figs. \ref{fig:6}-\ref{fig:9} the \ac{PDF} of the received signal amplitude, $R$, (calculated from the expressions for the statistics of $W$ by applying the corresponding change of variables) and the outage probability over \ac{IG}/\ac{TWDP} fading, contrasting all the results with \ac{MC} simulations. 

Figs. \ref{fig:6}-\ref{fig:7} depict the \ac{PDF} of $R$ for different values of the \ac{TWDP} parameters ($K$ and $\Delta$) and $m$. Note that, in order for these \acp{PDF} not to be overlapped, we set different values of $\overline{W}$ for each situation. Besides, to double-check the validity of the derived results, the theoretical plots have been calculated using the general expression in \eqref{eq:PDFTWDP_general} and the mixture representation in \eqref{eq:IGTWDP_mixturePDF} for Figs. \ref{fig:6} and \ref{fig:7}, respectively, showing all of them a perfect agreement with \ac{MC} simulations. 

As expected, with independence of the parameters of the \ac{TWDP} model, smaller values of $m$ render more sparse \acp{PDF}, since shadowing can be seen as an increment in the variance of the fast fading model. This is an important difference between composite models and \ac{LoS} shadowing models; while in the former the effect of increasing the shadowing severity is independent of the parameters of the baseline fading distribution, in the case of \ac{LoS} shadowing its impact depends on the power of this specular component, as showed in the analysis of, e.g., the $\kappa$-$\mu$ shadowed and the Fluctuating Beckmann models \cite{Paris2014, Ramirez2018}. On the other hand, as shown in Figs. \ref{fig:6}-\ref{fig:7}, larger values of $m$ reduce the shadowing severity and, in the limit ($m\to\infty$), the composite model converges to the baseline fading model.

The impact of assuming integer values of $m$ is represented in Fig. \ref{fig:8}, which represents the \ac{PDF} of the \ac{IG}/\ac{TWDP} for different values of $m\in\mathbb{R}^+$ and their corresponding integer values, obtained by rounding. We can extract a similar conclusion as in Section \ref{sec:Fitting}: as the value of $m$ increases, the impact of restricting it to be a positive integer becomes negligible. This reinforces the idea of considering $m\in\mathbb{N}^+$ for mild and moderate shadowing, since the reduced mathematical complexity that is achieved does not come at the price of a degraded accuracy in the fitting to empirical data measurements. 

Finally, Fig. \ref{fig:9} plots the outage probability over \ac{IG}/\ac{TWDP} fading in terms of the normalized threshold. It can be observed that the asymptotic expression in \eqref{eq:IGTWDP_asym} perfectly fits the theoretical curves (computed using \eqref{eq:IGTWDP_mixtureCDF} with $N=20$ terms computed) as $\gamma_{\textrm{th}}/\overline{\gamma}\rightarrow 0$. Interestingly, the severity of the shadowing (i.e., the value of $m$) only translates into a shift in the curves, as predicted from the theoretical analysis. Thus, the diversity order remains the same with independence of $m$.

\section{Conclusions}
\label{sec:Conclusions}

This paper provides a two-fold contribution in the context of composite fading models. On the one hand, we have performed, for first time in the literature, a thorough empirical validation of the \ac{IG} distribution based on data measurements from a wide variety of scenarios, showing that the use of the \ac{IG} distribution to model shadowing is well-justified. On the other hand, we have introduced a general methodology to characterize \ac{IG}-based composite models, which in many cases can be carried out by directly leveraging existing results in the literature. 

Specifically, we have proved that the \ac{PDF} and \ac{CDF} of the composite model can be expressed in terms of the \ac{GMGF} of the baseline fading distribution, a Laplace-domain statistic that is known for a wide variety of fading models. We have also provided simplified expressions in the case of considering the \ac{IG} distribution with integer shape parameter, showing how the impact of this assumption is negligible when the shadowing variance takes low or moderate values. Besides, we also proved that whenever the underlying fading model admits a mixture of gammas representation, then its composite version is directly given by a mixture of $\mathcal{F}$-distributions, allowing to leverage all the existing results for the latter model in the literature. Overall, we have shown that the use of \ac{IG} shadowing is not only justified from an empirical viewpoint, but also that it relaxes the mathematical complexity traditionally associated with composite models.   

In addition, the outage probability for \ac{IG}-based composite models is derived, proving that the shadowing severity does not affect the diversity order of the distribution, which is inherited from the underlying fading model.

Finally, as a direct application of our results, we have introduced two families of composite fading models based on $\kappa$-$\mu$ shadowed and \ac{TWDP} fading. These models extend most popular small-scale fading distributions in the literature to the composite case, giving different analytical expressions for their chief probability functions when \ac{IG} shadowing is considered. 

\vspace{-2mm}
\bibliographystyle{IEEEtran}
\bibliography{BibliografiaComposite}

\end{document}